\documentclass[a4paper,prb,showpacs,superscriptaddress]{revtex4}    
\usepackage{epsfig}

\begin{document}

\title{Influence of Grain size on the Electrical Properties of ${\rm
Sb_2Te_3}$ Polycrystalline Films.}

\author{P. Arun}
 \affiliation{Department of Physics \& Astrophysics, University of
Delhi, Delhi - 110 007, India}
  \email{arunp92@physics.du.ac.in, arunp92@yahoo.co.in, agni@physics.du.ac.in}
\author{A.G. Vedeshwar}
 \affiliation{Department of Physics \& Astrophysics, University of
Delhi, Delhi - 110 007, India}

\begin{abstract}
Resistance of vacuum deposited ${\rm Sb_2Te_3}$ films of thickness between 
100-500nm has been measured in vacuum. It is found that the resistance of the 
polycrystalline films strongly depends on the grain size and inter-granular 
voids. The charge carrier are shown to cross this high resistivity inter-
granular void by ohmic conduction. The barrier height as well as temperature 
coefficient of resistance are also shown to depend on the grain size and inter-
grain voids.
\end{abstract}


\maketitle

\section{Introduction} 
The transport mechanism and in turn the cause of resistance is of fundamental 
importance. Various model, especially for thin films, exist to understand the 
contribution from different scattering  mechanisms. The film resistance, 
however, may be due to a combination of three mechanisms,  namely (i) due to 
scattering from phonons, impurities and point defects etc., (ii) from film 
surface and (iii) due to grain boundaries which would be predominant in 
polycrystalline films\cite{1}. Different models exist to explain the dependence of 
resistance on film thickness. In the case of  scattering from the film surface 
the variation from film resistivity with film thickness was given by the Fuchs-
Sondheimer (F-S) relationship\cite{2}
\begin{eqnarray}
\rho (d) = \rho_o \left [ 1+ { l_g \over d} (1-p) \right ] \nonumber
\end{eqnarray}
Where ${\rm l_g}$ is the mean free path, 'd' is the film thickness and ${\rm
\rho_o}$ the resistivity of the bulk sample. The constant 'p' indicates the 
fraction of electrons being reflected from the surface. The value indicates the 
scattering mechanism, for example p=1 indicates specular reflection. A similar 
relationship was established by Mayadas and Shatzkes (M-S) \cite{3} to   
explain the scattering from grain boundaries, with a very similar functional
dependence with film thickness. However, the model is limited to very thin 
films with an added restriction that the grain size are of the order of film 
thickness. The grain boundary is defined as region between two grains where 
crystal orientation changes. The transport properties of ${\rm Sb_2Te_3}$ films 
like resistivity, Hall coefficient, mobility and Seeback coefficient have been 
extensively studied \cite{4, 5, 6, 7, 8}, and reports in literature indicate 
films to be p-type with narrow band gap. However, all these reports are silent 
on the mechanism of scattering and in turn the source of resistivity. Only
Damodara Das et al \cite{7}  have reported resistivity as a function of 
thickness ${\rm (50nm < d <120nm)}$. The article states average grain size to 
be of the order of film thickness and indicates the scattering
mechanism to be that of grain boundary scattering in accordance to the M-S
model. However, no report exists on the variation of resistance or resistivity 
of ${\rm Sb_2Te_3}$ films with grain size in thicker films. This article 
investigates variation of resistance in films whose
thickness is enough to assume that the variation in resistivity is
independent of defects and specular scattering.

\section{Experimental}

Films of ${\rm Sb_2Te_3}$ were grown on glass substrates kept at room 
temperature, using thermal evaporation method. ${\rm Sb_2Te_3}$ ingot of high 
purity (99.99\%) supplied by Aldrich (USA) were used as the starting material. 
The crushed ingot were evaporated from molybdenum boat at a vacuum better than 
${\rm 10^{-6}}$Torr. The film thickness was measured using Dektek IIA surface 
profiler. The movement of the mechanical stylus across the edge
of the film determines the step height or the films thickness. Indium
contacts were grown on the glass substrates before they were placed in the 
chamber, such that a strip of ${\rm Sb_2Te_3}$ film of dimensions ${2.3cm
\times 1.65mm}$ could be fabricated on these contacts using a mask. 
The I-V characteristics of the films were measured by four probe method. It
was found to be linear between 25mV-24V, showing the ohmic nature of indium 
contacts as well as the polycrystalline film for applied field. The films' 
resistance were measured by an high input impedance digital multimeter. The 
structural and compositional analysis of these films were done using Phillips 
PW1840 X-ray diffractometer and Shimadzu ESCA750 (Electron Spectroscopy for 
Chemical Analysis). The films were found to be stoichiometrically     
uniform over the area 5cm x 5cm as determined by ESCA carried out in various
regions of the film. The morphological analysis was done with JOEL 840 Scanning
Electron Microscope (SEM). The as grown films showed tendency of ageing
\cite{9}, where the resistance of the film varied with time and saturated to a 
constant value in couple of weeks. The results
presented in this article are of films which had achieved such saturation.

\section{Results and Discussion}

\subsection{Variation Resistance with grain size}

The average grain size was determined from both SEM micrographs and X-ray
diffractograms. The grain size was calculated using the Full Width at Half 
Maxima (FWHM) of X-Ray peaks \cite{10}. The results of grain size found by both 
methods were in agreement. A plot between the film thickness and grain size 
shows no trend (fig 1).  This variation in grain size with film thickness may 
be a result of not having perfectly identical conditions during film evaporation. 
It also represents the randomness of the growth process. This shows that the 
average grain size is not proportional to the film thickness
and resistance or resistivity will have to be studied both as a function of
thickness and grain size to resolve the main contributor in scattering 
mechanism. The F-S theory shows that the contributions from the surface leads 
to an inverse proportionality with thickness (equation 1), where the model is 
restricted to cases where the charge carriers mean free path is of the order of 
the film thickness ${\rm (l_g \sim d)}$. However, since the samples in our study 
have thickness between 130-500nm, the film thickness is far 
greater then the mean free path. Beyond this limit one can assume the films
resistivity to be same as that of the bulk, showing no further change with 
increasing film thickness. Thus, film's resistance in this limit should only 
fall inversely with thickness. Figure 2 shows the resistance of aged 
${\rm Sb_2Te_3}$ films falling linearly with increasing thickness. It can
be understood trivially, that for the resistance of the film to vary
linearly with thickness as shown in figure 2, the resistivity would have to 
show a parabolic relationship with film thickness.
Another important contributor to resistance is the grain boundary. However,
that too requires an inverse proportionality with thickness. This lack of trend
may be due to the assumption in M-S theory that the grain size is proportional 
to the film thickness, which is not the case here. It is clear that in the 
present study the surface scattering and grain boundary scattering do not 
contribute to the film resistance. Hence, to investigate the influence of 
the grain size on transport properties, variation of resistance with 
grain size was studied. Figure 3 shows the variation of film
resistance with grain size. As stated earlier the average grain size was
determined from both SEM micrographs and X-ray diffractograms.
The grain boundary is defined as region between two grains where crystal
orientation changes. The representative micrographs of ${\rm Sb_2Te_3}$ in 
figure 4 however, show large distances between two grains. The grains tend 
to have the resistivity of the bulk, however, even if there is an 
inter-connectivity between two neighbouring grains this    
region will have high resistivity by purely geometry of narrowing \cite{11}.
These voids, hence would definitely contribute differently from the defined 
grain boundary in M-S theory.    

Volger's model \cite{12} assumes the film to be made up of cubical grains of 
edge size 'a'  arranged in an ordered manner, as shown in fig. 5a, with equal 
spacing between the neighbouring grains. The inter-grain distances are 
different along x, y and z directions and are same along any one direction. 
Consider the film has 'q' number of grains arranged regularly at equal inter-
grain spacing ${\rm 't_x'}$ along the length 'l' and 'r' and 'p' grains arranged 
along the width and thickness of the film. Also, the resistance is measured 
along the length of the film by taking the contacts across the cross-section 
in the yz plane, then the points A-B, C-D etc. shown in fig. 5a are at equal 
potential. The equivalent dc circuit of this arrangement of measurement would 
be as shown in figure 5b, where ${\rm 'R_b'}$ represents the high resistance of 
the inter-grain voids \cite{13, 14, 15}.  As can be seen in figure 5b,  the    
whole film can be considered to be a parallel combination of 'pr' resistive
elements, where resistance of each element is given by \cite{11, 16, 17}
\begin{eqnarray}
R_1 = qR_g + (q-1)R_b\nonumber
\end{eqnarray}
Thus, the net resistance along the length of the film between the two
contacts would be given as
\begin{eqnarray}
R_{net} =  {qR_g + (q-1)R_b \over pr}\nonumber
\end{eqnarray}

Seto \cite{18} made a similar simplification step by assuming the problem to be
that of one dimension. ${\rm 'R_b'}$, the high resistance of the inter-grain 
voids, is a function of ${\rm 't_x'}$ which in turn would depend on the 
mechanism by which charge carriers would cross the inter-grain boundary. Many 
suggestions have been made for explaining the cross over, such as
ohmic conduction, tunnelling or thermionic emission \cite{19}. It may also be a
combination of these, depending on the actual inter-grain distances. The 
resistance of such a film, assuming ohmic conduction in between grains is
given as \cite{20}
\begin{eqnarray}
R_{net} = \alpha {1+ kx \over (1+x)^2} 
\end{eqnarray}
where ${\rm \alpha}$ is a proportionality constant, given as 
\begin{eqnarray}
\alpha = {\rho_g l^2 \over V-V_{void}} \nonumber 
\end{eqnarray}
In true sense ${\rm \alpha}$ is not a constant since ${\rm V_{void}}$ will 
depend on the grain size, as also film dimensions, including it's thickness. 
However, ${\rm V_{void}}$ is assumed to be a slow varying function of film 
thickness, or constant. The constant 'k' 
represents a ratio of the inter-grain region's resistivity to the grain's 
resistivity. Since the void resistivity is large, 'k' obviously is a very 
large entity. The variable  is a ratio of the inter-grain length and  
the grain edge or 
\begin{eqnarray}
x =  {t_x \over a}\nonumber
\end{eqnarray}
where 'a' is the grain size, assuming as in Volger's model, the grains to be
cubic in nature. The inter-grain distance, ${\rm 't_x'}$ is extent of void in 
'x' direction (along length of the film strip), since the resistance is 
measured along the length of the 
film. The inter-grain distance varies as a function of the grain size depending 
the mechanism of grain growth. Since, the ${\rm Sb_2Te_3}$ films aged to a 
hexagonal crystal state, with ${\rm c >> a}$, it should show easier
grain growth along the length and width as compared with that along restrictive 
film thickness. The films hence aged with the c-axis aligned normal to the 
substrate plane \cite{20}. As per Volger's model 'q', the grain number along 
the length, would be decreasing more rapidly than 'p', that along the film's 
thickness, leading to a general trend of decrease in resistance. Thus, the 
variation of inter-grain distance with grain size for
the films in consideration would be given as \cite{20}
\begin{eqnarray}
t_x = a \left ( {pra^2l - \Delta V \over \Delta V -pra^3} \right )
\end{eqnarray}
where ${\rm \Delta V= V-V_{void}}$. Thus, it can be seen that the variable, 
is a function of the grain size, 'a'. The increase in void size with increasing 
grain size can be appreciated from the representative SEM micrographs. Equation 
2, is physically valid for positive values, which requires
\begin{eqnarray}
pra^2l > \Delta V > pra^3\nonumber
\end{eqnarray}

Considering a extreme case of ${\rm pra^2l >> \Delta V >> pra^3}$ along with 
the stated assumption that ${\rm \Delta V}$ is constant, then equation 2 maybe 
written as
\begin{eqnarray}
t_x = a \left ( {pra^2l\over \Delta V} \right )\nonumber
\end{eqnarray}
The variable 'x' required for equation (1) can then be expressed as
\begin{eqnarray}
x = {t_x \over a} = a^2 \left ( {prl\over \Delta V} \right )= \beta a^2 
\end{eqnarray}

This increase in inter-grain distance with growing grain size was discussed
in our earlier work \cite{20}. Hence, using equation (3) the films resistance 
given by equation (1), can be expressed as
\begin{eqnarray}
R_{net} = \alpha {1+ k\beta a^2 \over (1+\beta a^2)^2} 
\end{eqnarray}
Equation (4) fits quite well to the experimental observations as shown by
the solid line in figure 4.  The values of the constants evaluated by fitting 
are ${\rm \alpha}$ =2308 ${\Omega}$, ${\rm \beta = 20.44 \times 10^{-6}
\AA^2}$ and ${k \sim 54}$. As stated earlier, the constant k is a ratio of the
high resistivity of the inter-grain region as to the low resistivity of the 
grains. The numerical value shows the resistivity of the inter-grain region 
will be nearly ${\rm 10^2}$ times that of the low resistivity    
region, which is consistent with the with the assumption that inter-grain
region can be assumed to be a path of high resistance.

\subsection{Variation of Barrier Height with grain size}

The voids between neighbouring grains would present itself as a barrier
which the charge carriers would have to transverse to establish current flow. 
The magnitude of the barrier height can be computed from the slope of the plot 
between ${\rm ln(\sigma)}$ and temperature inverse (1/T in Kelvin). The barrier 
height was calculated using this method for various film thickness. The 
variation is shown in figure 6. The variation in barrier height with film
thickness maybe due to one or a cumulative effect of the following (i)
variation in the grain size of the polycrystalline film, (ii) a large density 
of dislocations, (iii) quantum size effects and (iv) change in film 
stoichiometry. Since the film thickness of this study is large  
the quantum size effect is immediately ruled out. Careful growth technique
followed by ageing would minimise the contribution due to dislocation and off
stoichiometric compositions, however, can not be completely ruled out. The 
major contribution hence would be due to the size of the grains. Slater
\cite{21} estimated the barrier height as a function of grain size by modelling 
grain boundary as a pn type of structure. The variation is given as
\begin{eqnarray}
E_b = E_o + {N_oe^2 \over 4k\epsilon_o}\left (t_x -{N \over N_o}a \right )^2
\end{eqnarray}
where ${\rm N_o}$ is the doping concentration, N the carrier concentration, k 
the dielectric constant of the material. The barrier height increases with 
grain size for ${\rm Na/N_o > t_x}$, which would be the case in pure samples 
${\rm (N>N_o)}$. A fit for the experimental data using equation 3, which states 
that the 
barrier width in proportional to the square of the grain size, along with the
estimated proportionality constant (b) and equation 5 is shown by the continuous 
line in figure 6. The fit shows reasonable agreement, however it does indicate
possible contributions from dislocations etc. The values of the coefficients of
equation 5 are ${\rm E_o}$=5.66meV, ${\rm N/N_o}$ = 90 and ${\rm N_oe^2/4k
\epsilon_o = 2.65 \times 10^{-9}meV- \AA^2}$. The ratio of ${\rm N/N_o}$ may 
appear to be very small, however, it should be noted that the curve fitting was
done using the earlier estimated proportionality constant (${\rm \beta}$). Thus, 
the influence of the inter-grain voids and grain size on the magnitude of 
barrier height and various characterising parameters of  
the film is evident. This though expected is different from an earlier study
by Rajagopalan et al \cite{4} on films of ${\rm Sb_2Te_3}$ with thickness 
between 160nm and 800nm which reported that     
the barrier height was independent of the film thickness. In the next
section we investigate the role of the inter-grain voids on another parameter 
used to characterise the transport properties of a material. 

\subsection{Temperature coefficient of Resistance}

The resistance of the film is a function of temperature. The variation of 
resistance with temperature in general is expressed as
\begin{eqnarray}
R = R_o(1+ {\alpha \over R_o} T)\nonumber
\end{eqnarray}
Thus the temperature coefficient of resistance or TCR \cite{22} is given as
\begin{eqnarray}
{1 \over R_o}{dR \over dT}  = {\alpha \over R_o}
\end{eqnarray}
While TCR is positive for metals, it is negative in case of semiconductors. 
The F-S model for very thin films states that the variation of TCR with 
film thickness follows an identical form as expressed by equation (1). However, 
there seems to be no model in the literature to explain the variation of TCR 
with either film thickness or grain size for films with thickness greater then 
the mean free path of their charge carriers. A plot between TCR and grain size 
seems scattered (not shown). However, the plot between TCR and barrier 
height is a straight line, figure 7, with slope 
${\rm -8.6 \times 10^{-5} (^o C-meV)^{-1}}$ and intercept 
${\rm -2.66 \times 10^{-5} (^oC^{-1})}$. It 
immediately follows from the linearity between TCR and ${\rm E_b}$ along with 
equation (5) that the TCR would be a polynomial function of the grain size.
This explains the seemingly scattered data points of TCR with grain size as 
discussed. It also explains the lack of any model or theory on the variation 
of TCR with grain size. The relationship shows that with increasing barrier 
height, the rate of change of resistance with temperature (dR/dT or TCR, 
equation 6) becomes increasingly smaller. The negative temperature 
coefficient  of resistance in semiconductors is a result of increasing charge
carriers due to breaking covalent bonds. An increased barrier height implies 
its more difficult for the   
charge carriers to escape into the voids from the grains, confining the
increased number of
carriers inside the grain itself. This rapidly brings down the resistance of
the grain contributing to a negative TCR proportional to the barrier height.

\section{Conclusions}
The dc transport properties of ${\rm Sb_2Te_3}$ films with thickness between 
130-500nm
have been discussed. The properties showed no size effects as was expected,
since the film thickness was far greater then the charge carriers mean free 
path. The films resistance, the barrier height and temperature coefficient of 
resistance (TCR) showed a strong dependence on both the grain size and inter-
grain void. The inter-grain void was approximated to vary with increasing grain 
size, enabling to study the above properties of
the films as a function of  grain size.

\begin{acknowledgments}
The contributions of Rajni Jain, Pankaj Tyagi and Naveen Gaur are
acknowledged.

\end{acknowledgments}

\pagebreak

\pagebreak

%%%%%%%%%%%%%%%%%%%%%%%%%%%%%%%%%%%%%%%%%%%%%%
%  Figure inclusion (at the end of paper)
%%%%%%%%%%%%%%%%%%%%%%%%%%%%%%%%%%%%%%%%%%%%%%
\begin{figure}
\begin{center}
\epsfig{file=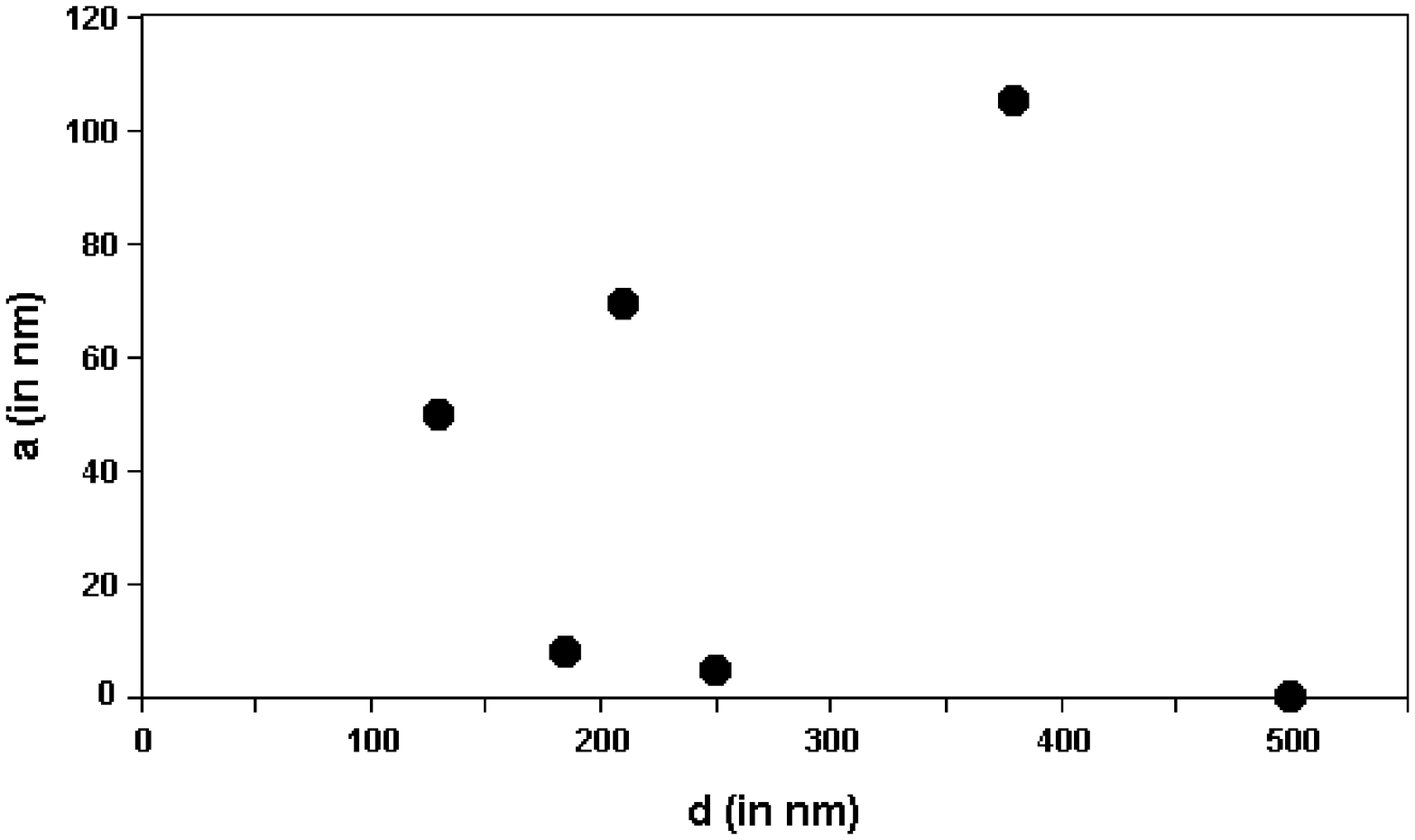,width=5 in}
\caption{ The variation of grain size in the polycrystalline films for
various thickness of ${\rm Sb_2Te_3}$ grown by thermal evaporation method, 
show no trend.} 
\label{fig:1}
\end{center}
\end{figure}

\begin{figure}
\begin{center}
\epsfig{file=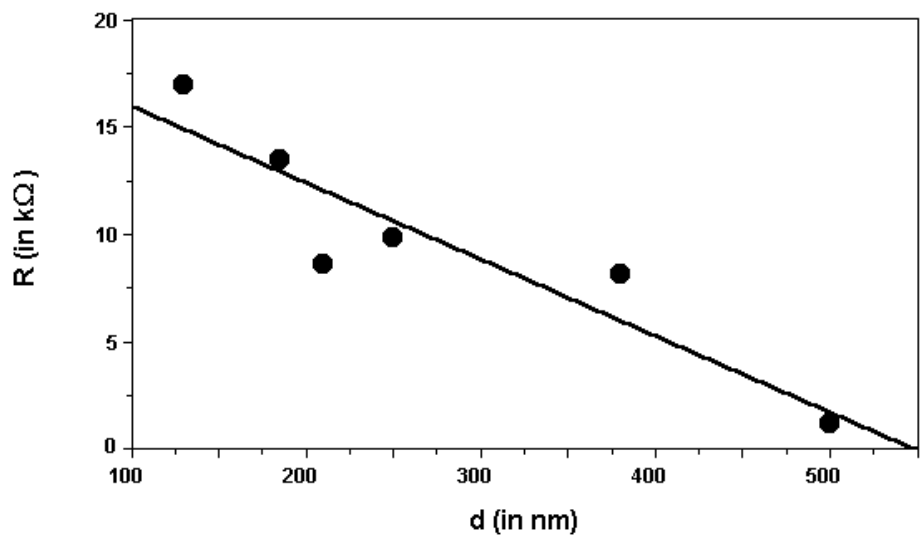,width=5 in}
\caption{ Variation of  film resistance with film thickness of various
polycrystalline ${\rm Sb_2Te_3}$ films after they have completely aged.} 
\label{fig:2}
\end{center}
\end{figure}

\begin{figure}
\begin{center}
\epsfig{file=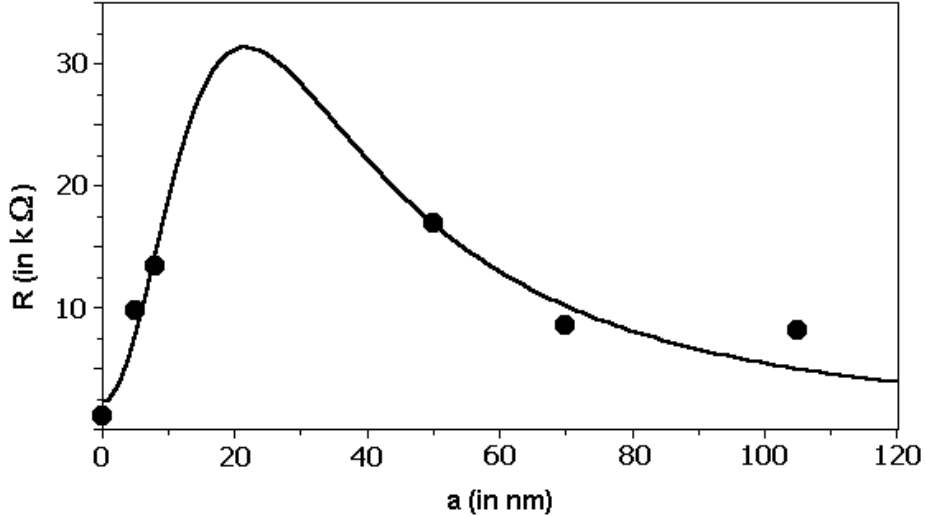,width=5 in}
\caption{ Variation of  film resistance with grain size of various 
polycrystalline ${\rm Sb_2Te_3}$ films. The continuous curve is a fit of the 
experimental points using equation (4).}
\label{fig:3}
\end{center}
\end{figure}

\begin{figure}
\begin{center}
\epsfig{file=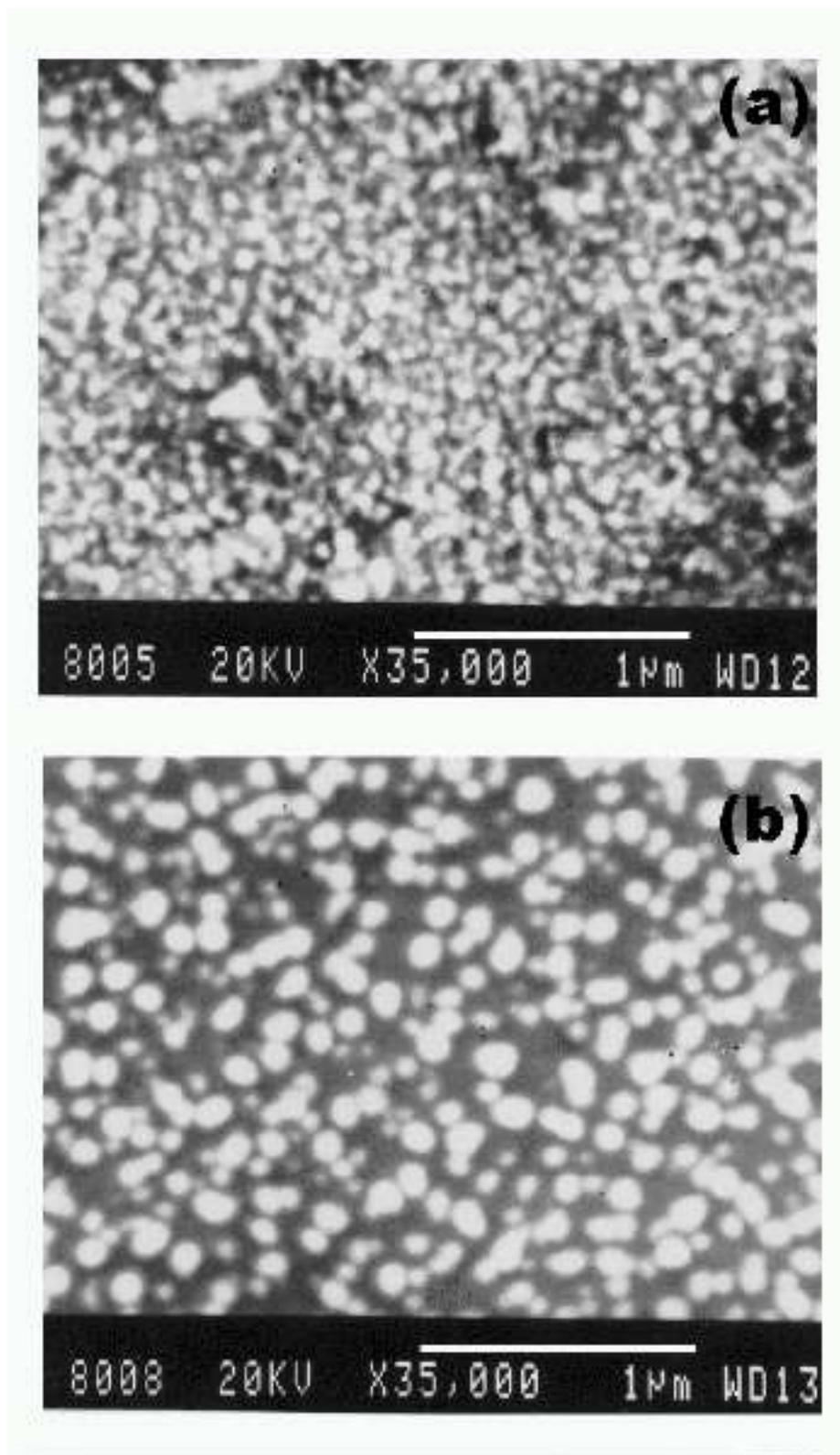}
\caption{ SEM micrographs of two films with different thickness (a) 130nm
and (b) 380nm, showing grains with voids between neighbouring grains. It is 
evident that as grain size increases the voids also increase.}
\label{fig:4}
\end{center}
\end{figure}

\begin{figure}
\begin{center}
\epsfig{file=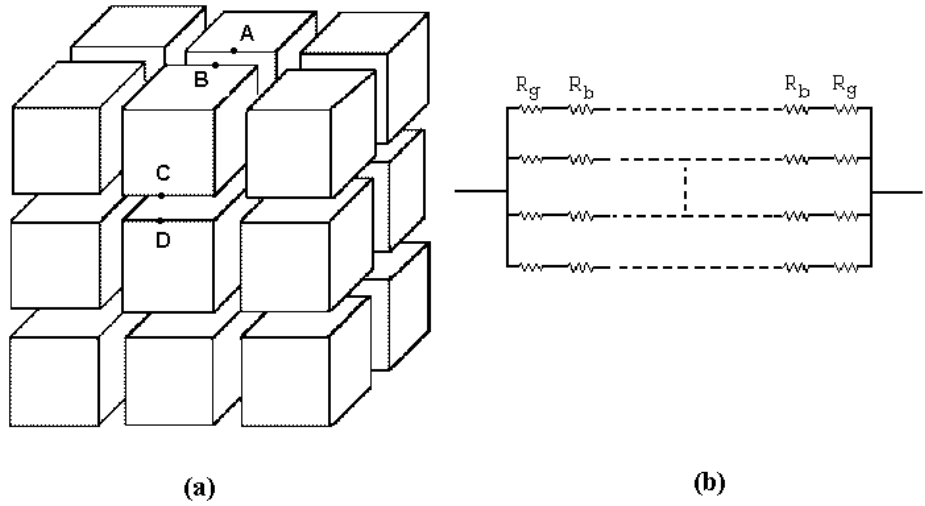,width=5 in}
\caption{  Figure shows (a) an idealised assumption of how cubic grains are
arranged along the dimensions of a polycrystalline film and (b) shows an 
equivalent circuit of the a polycrystalline film based on simplified 
assumptions.}
\label{fig:5}
\end{center}
\end{figure}

\begin{figure}
\begin{center}
\epsfig{file=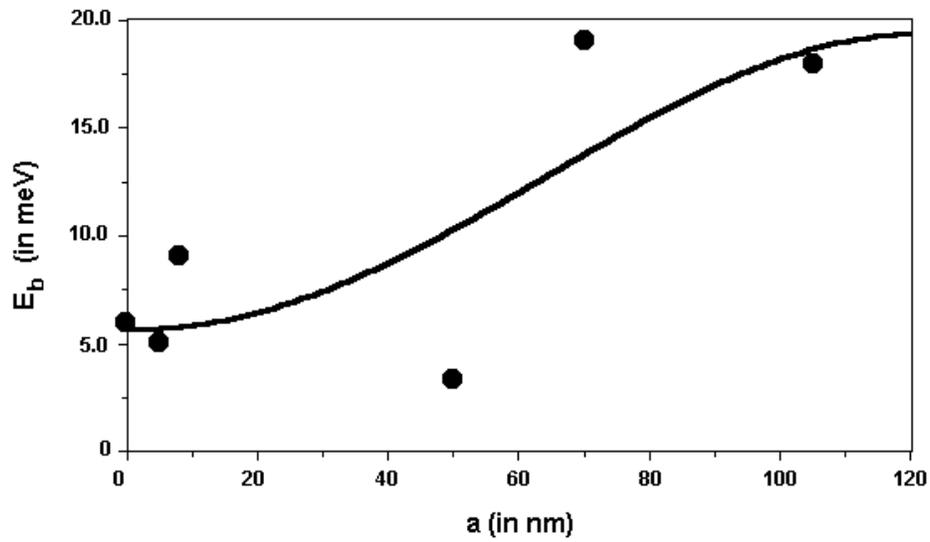,width=5 in}
\caption{ Plot shows the variation of energy barrier height with improving
grain size. The continuous curve is a fit of the experimental point using 
Slater's model (equation 5). The scattered points indicate other influences 
also on the barrier height. }
\label{fig:6}
\end{center}
\end{figure}

\begin{figure}
\begin{center}
\epsfig{file=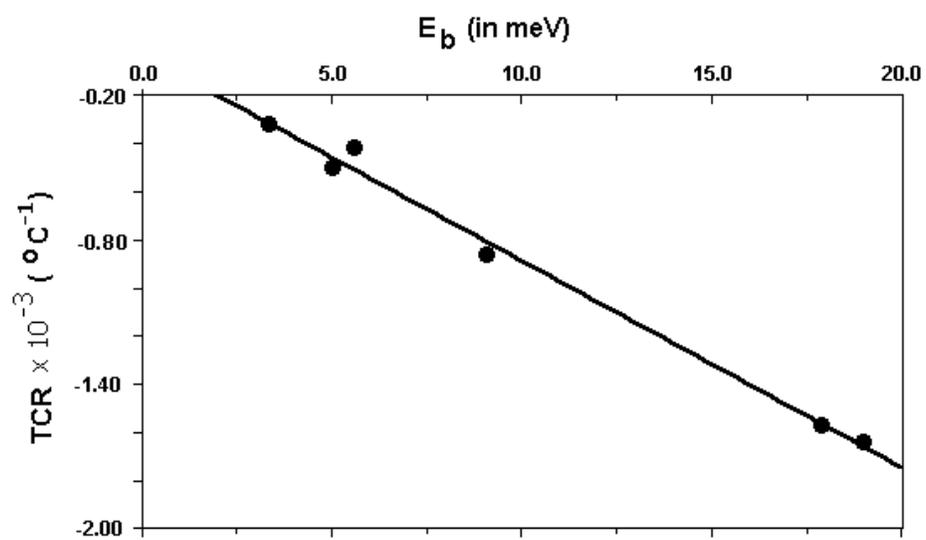,width=5 in}
\caption{ The linear variation of temperature coefficient of resistance
(TCR) with the barrier height implies the dependence of with the grain size.}
\label{fig:7}
\end{center}
\end{figure}
\end{document}